\documentclass[aps,prl,twocolumn,nofootinbib]{revtex4-1}

\RequirePackage{graphicx}
\usepackage{caption}
\usepackage{subcaption}
\newcommand{\bea}{\begin{eqnarray}}
\newcommand{\eea}{\end{eqnarray}}
\usepackage{multirow}
\usepackage{verbatim}
\usepackage{xcolor}

\newcommand{\Jpsi}{J\! /\! \psi}
\newcommand{\be}{\begin{eqnarray}}
\newcommand{\ee}{\end{eqnarray}}

\usepackage{amsmath}	
\usepackage{booktabs}
\begin{document}

\title{
Rich structure of the hidden-charm pentaquarks near threshold regions
}

\author{Alessandro~Giachino$^{1,2}$}
\email[]{alessandro.giachino@ifj.edu.pl}
\author{Atsushi~Hosaka$^{3,4}$}
\email[]{hosaka@rcnp.osaka-u.ac.jp}
\author{Elena~Santopinto$^{1}$}
\email[Corresponding author: ]{elena.santopinto@ge.infn.it}
\author{Sachiko~Takeuchi$^{3,5,6}$}
\email[]{s.takeuchi@jcsw.ac.jp}
\author{Makoto~Takizawa$^{5,7,8}$}
\email[]{takizawa@ac.shoyaku.ac.jp}
\author{Yasuhiro~Yamaguchi$^{5,9}$}
\email[]{yamaguchi@hken.phys.nagoya-u.ac.jp}

\affiliation{$^1$Istituto Nazionale di Fisica Nucleare (INFN), Sezione di
Genova, via Dodecaneso 33, 16146 Genova, Italy
}
\affiliation{$^2$Institute of Nuclear Physics Polish Academy of Sciences Radzikowskiego 152, 31-342 Cracow, Poland
}
\affiliation{$^3$Research Center for Nuclear Physics (RCNP), Osaka
University, Ibaraki, Osaka 567-0047, Japan}
\affiliation{$^4$Advanced Science Research Center, Japan Atomic Energy Agency, Tokai, Ibaraki 319-1195, Japan}

\affiliation{$^5$
Meson Science Laboratory, Cluster for Pioneering Research, RIKEN, Hirosawa, Wako, Saitama 351-0198, Japan
}

\affiliation{$^6$Japan College of Social Work, Kiyose, Tokyo 204-8555, Japan}

\affiliation{$^7$Showa Pharmaceutical University, Machida, Tokyo
194-8543, Japan}

\affiliation{$^8$J-PARC Branch, KEK Theory Center, Institute for Particle and Nuclear Studies, KEK, Tokai, Ibaraki 319-1106, Japan}

\affiliation{$^{9}$Department of Physics, Nagoya University, Nagoya 464-8602, Japan}

\begin{abstract}
The recent abundant observations of pentaquarks and tetraquarks by high-energy accelerator facilities
indicate the realization of the conjecture by Gell-Mann and Zweig, and by De~Rujula, Georgi 
and Glashow~\cite{Gell-Mann:1964ewy,Zweig:1964ruk,DeRujula:1976ugc}.  
We construct a coupled-channel model 
for the hidden-charm pentaquarks  with strangeness whose quark content is $udsc \bar c$, 
$P_{cs}$, described as $ \Lambda_c \bar{D}_s^{(*)}, \Xi_c^{('*)} \bar{D}^{(*)}$  
molecules coupled to the five-quark states.  These molecules are formed by the suitable cooperation of heavy quark and chiral symmetries.  
We reproduce the  experimental  mass and quantum numbers  $J^P$ of  $P_{cs}(4338)$ for which LHCb has just  announced the discovery. We make other predictions for new $P_{cs}$ states as molecular states near threshold regions that can be studied by LHCb. 
\end{abstract}

\maketitle

The past decade has witnessed tremendous progress in the experimental and theoretical explorations of the exotic hadrons.
These are the strongly-interacting particles made up of quarks, but are considered to have more complicated structures than those of ordinary hadrons such as protons and neutrons, 
which were already mentioned in the early stages of the prediction of quarks and the discovery of the charm quark~\cite{Gell-Mann:1964ewy,Zweig:1964ruk,DeRujula:1976ugc}. The saga of exotic hadrons dates back to 2003,  when the Belle Collaboration discovered the first tetraquark candidate,  $X(3872)$, with quark content
$\bar c c \bar u u$~\cite{Belle:2003nnu}.
While further analyses are going on, the  Large Hadron Collider beauty (LHCb) experiment  revealed   as many as 59 signals for new hadrons~\cite{LHCbpentas2}.
What should we learn from the observation? The question that should be clarified was nicely formulated in \cite{Brambilla:2022ura}: 
how are quarks organized inside these multiquark states $-$ as compact objects with all quarks within one confinement volume, 
interacting via color forces,  or as deuteron-like hadronic molecules, bound by light-meson exchange? 
Indeed, though the existence of these states has now been confirmed, their internal structure is still controversial.

A new phase of quest was triggered in 2015, when the LHCb collaboration reported the first discovery of two pentaquark states, which have been called $P^+_c(4380)$ and $P^+_c(4450)$~\cite{Aaij:2015tga}, in  $\Lambda_b^0 \to P_c^{+} K^{-} \to (\Jpsi p) K^{-}$ decay channel.  
The quark content of these states is implied by the observed particles $\Jpsi p \sim \bar cc uud$.  
Four years later, a new analysis \cite{Aaij:2019vzc} with nine times more statistics was performed; this revealed  $P^+_c(4312)$, as well as the splitting of the 
$P^+_c(4450)$ into two narrow peaks, $P^+_c (4440)$ and $P^+_c(4457)$.  
Later on, evidence emerged for  a  new pentaquark state with mass  
$M\simeq 4337$ MeV~\cite{LHCb:2021chn}.  
All of the above  states appear near a two-hadron threshold, for instance $P^+_c(4312)$  near the threshold of  the $\bar D$ meson and  $\Sigma_c$ baryon.  

This was not the end of story.  
In 2020, the first evidence of a pentaquark with strangeness, $P_{cs}(4459)$, was reported in the  $\Xi_b^{-}\rightarrow P_{cs}(4459) K^-\to  (\Jpsi \Lambda) K^- $ decay  channel with statistical significance of 3.1 $\sigma$ \cite{Aaij:2020gdg}.  
It is worth noting that this resonance can be equally well described by a two-peak structure, 
with the two peaks split by 13 MeV: $P_{cs}(4455)$ and  $P_{cs}(4468)$~ \cite{Aaij:2020gdg,Karliner:2022erb}.  
The experimental masses of $P_{cs}(4455)$ and  $P_{cs}(4468)$ are $M=4454.9 \pm 2.7$ MeV
and $M=4467.8 \pm 3.7$ MeV.   
According to LHCb, the two-peak structure hypothesis has the same statistical significance as the single-peak hypothesis~\cite{Aaij:2020gdg}.  
Unfortunately, owing to limited signal yield, the $J^P$ of the $P_{cs}(4455)$ and  $P_{cs}(4468)$ states could not be determined in this analysis~\cite{Aaij:2020gdg}. 
$P_{cs}(4455)$ and  $P_{cs}(4468)$ lie below the $\Xi_c \bar{D}^{*}$ threshold and so this situation is similar to what happens in the non-strange sector to the two states  $P_c(4440)$ and $P_c(4457)$, which are just below the $\Sigma_c\bar{D}^{*}$ threshold.

Very recently LHCb has announced the discovery of a new state
 with  mass $M=4338.2 \pm 0.7$ MeV and width $\Gamma=7.0 \pm 1.2$ MeV  with statistical significance $> 10 \, \sigma$ in $B^{-}\to P_{cs} \bar{p} \to (\Jpsi \Lambda) \bar{p}$ : thus, $P_{cs}(4338)$~\cite{LHCbpentas2}. The amplitude analysis performed by LHCb favors  spin and parity $J^P=\frac{1}{2}^{-}$ \cite{LHCbpentas2}.
Again, these states appear very close to  a two-hadron threshold.  
Indeed, this applies not only to the pentaquarks but also to tetraquarks, 
well-known candidates for which are 
$X(3872)$ and $T_{cc}(3875)$
~\cite{LHCb:2021vvq,LHCb:2021auc}.

  In Ref. \cite{Xiao:2019gjd}  several hidden charm  pentaquarks  with strangeness have been predicted by means of a $SU(4)$ extension of the Local Hidden Gauge approach
   to the charm sector and, in particular,  a $J^P=\frac{1}{2}^-$ state with  mass $4277$ MeV; in Ref. \cite{Wang:2019nvm} a pentaquark state with $J^P=\frac{1}{2}^-$ and  mass $4319.4^{+2.8}_{-3.0}$ MeV has been predicted  within chiral effective field theory with only leading-order contact interactions;
 in Ref. \cite{Chen:2020uif} a pentaquark state with  $J^P=\frac{1}{2}^-$ and mass $4290^{+130}_{-120}$ MeV  has been predicted with QCD sum rule; 
  in Ref. \cite{Ali:2019clg} a pentaquark state with $J^P=\frac{1}{2}^-$  and mass $4292$ MeV has been predicted  in the compact diquark model; in Ref. 
    \cite{Ferretti:2021zis} a pentaquark state with $J^P=\frac{1}{2}^-$  and mass $4485$ MeV has been predicted within the hadrocharmonium model.
Finally, in Ref.  \cite{Chen:2022onm}   a  coupled-channel calculation limited to  $\Xi_{c}^{(*')} \bar{D}^{(*)}$ channels has been studied as a function of 
the cut-off parameter.
Just after the LHCb collaboration  announcement of  5$^{\text{th}}$ of June \cite{LHCbpentas2}  a quark model interpretation appeared  \cite{Karliner:2022erb},
and in Ref. \cite{Wang:2022mxy} a molecular interpretation within a coupled-channel model limited to $\Xi_{c}^{(')} \bar{D}^{(*)}$ channels, 
while in Ref. \cite{Yan:2022wuz}  within a coupled-channel model  using only contact range interactions.
  Finally,  in \cite{Burns:2022uha}  a triangle singularity interpretation is  proposed.

 The new $P_{cs}(4338)$ state is very intriguing because, as discussed by LHCb, its mass is very close to the $\Xi_c^{+}\bar{D}^{-}$ meson-baryon threshold, which lies at 2467.7 + 1869.7 = 4337.4 MeV, and indeed  its  favorable quantum numbers, $J^P =1/2^{-}$, are just what one expects for the $\Xi_c^{+}\bar{D}^{-}$ meson-baryon system in an S-wave. The most natural decay channel for such a state should be the $J/\Psi \Lambda$ channel, whose threshold is located 126 MeV below the $P_{cs}(4338)$ mass. The 7 MeV width $P_{cs}(4338)$ is unnaturally small for such a large phase space so there must be some decay-suppressing mechanism at work \cite{Karliner:2022erb}.  The small experimental decay width in  the $J/\Psi \Lambda$ channel can be understood only if the  $P_{cs}(4338)$ is a very shallow resonant $\Xi_c^{+}\bar{D}^{-}$ meson-baryon system, in which the formation of a $c\bar{c}$ pair is suppressed by the long distance between the $\bar{D}^{-}$ meson and the $\Xi_c^{+}$  baryon \cite{Karliner:2022erb}.

If two (or more) particles interact suitably with each other
they form a weakly bound or resonant state,
just as atoms form molecules.  
$P_{cs}$s are the molecules of $\Lambda_c$-$\bar{D}_s^{(*)}$ and  $\Xi_c^{(')}$-$\bar{D}^{(*)}$.  
The molecule's constituent hadrons, such as $\Xi_c^{(')}$ and $\bar{D}^{(*)}$, are colorless clusters of quarks 
that are strongly bound by colored force of Quantum Chromodynamics (QCD) mediated by gluons.
Otherwise, the constituent hadrons interact weakly via a colorless force mediated by mesons.  
Hence, the molecular states are realized by finely-tuned conditions 
of the formation of the constituent hadrons as quark clusters with suitable interaction strength and masses.
While the masses of constituent hadrons are well known, not much is known  about their interaction.
Recently, LHC has started to look for correlation functions in  high-energy hadron-hadron collisions~\cite{ALICE:2020mfd} and lattice QCD calculations~\cite{Hatsuda:2018nes}
are  ongoing.  
But these  have not yet been fully achieved.  

\begin{figure}[h]
  \centering
  \includegraphics[width=0.8\linewidth]{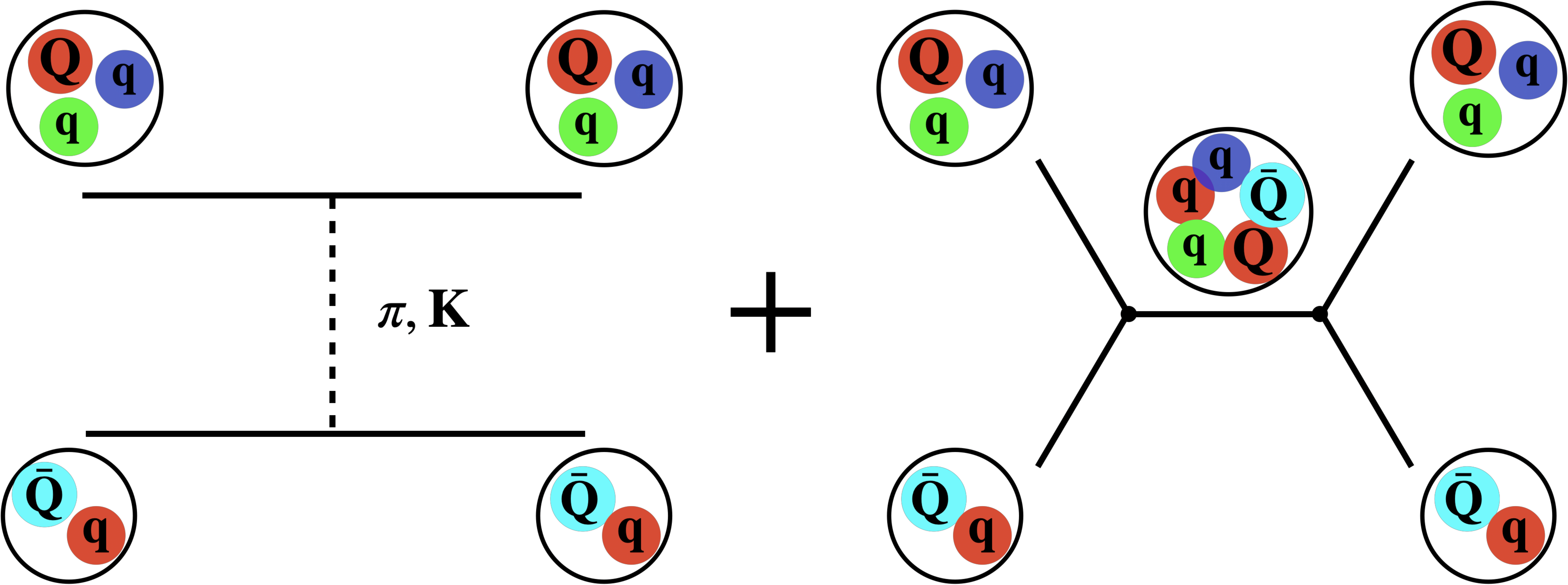}
\caption{Pictorial representation of the pentaquarks described as five-quark core + meson-baryon molecular components interacting via 
  $\pi$ and $K$ exchange-mediated potential.}
\label{figcoupling}
\end{figure}

Hence, we attempt a model construction on the basis of the  knowledge of the strong interaction
that  has been accumulated so far.   
The best established    model is the  one-pion and kaon exchange force between the constituent hadrons
via  meson coupling to light $u, d, s$ quarks.  
The interactions between the heavy baryons and heavy mesons via pion and kaon exchange are derived from Lagrangians that satisfy the heavy quark and
chiral symmetries~\cite{Yamaguchi:2019vea}.
This is illustrated on the left in Fig.~\ref{figcoupling}.  

If we consider that the hadron size is in the order of half fm and the distance between hadrons in a hadronic molecule 
(molecular size) is one fm or larger, then the hadrons may overlap for quite some time while forming the molecule.  
Unlike ordinary molecules, in which constituent atoms repel each other at short distances, constituent hadrons may not.   
In such a situation, there should exist a transition 
between hadronic molecular components and compact five-quark 
components.
Refs.~\cite{Santopinto:2016pkp,Takeuchi:2016ejt} studied 
the possible configurations of the compact five-quark components and their energies.
Thus, we arrive at a coupled-channel model of hadronic molecules 
and five-quark states, as shown in Fig.~\ref{figcoupling}~\cite{Yamaguchi:2017zmn,Yamaguchi:2019seo} 
 (see also the supplementary material). 
The coupling structure of the molecules and five-quarks is dictated by the so-called 
spectroscopic factor, which is familiar in the discussion of cluster structure of atomic nuclei. 
The spectroscopic factor corresponds to the probability of finding a molecular component in a five-quark state.
 Thus,  there is only one parameter for the overall coupling strength  which we have denoted as $f/f_0 \equiv F$.

\begin{figure}[h!]
  \centering
  \includegraphics[width=0.9\linewidth]{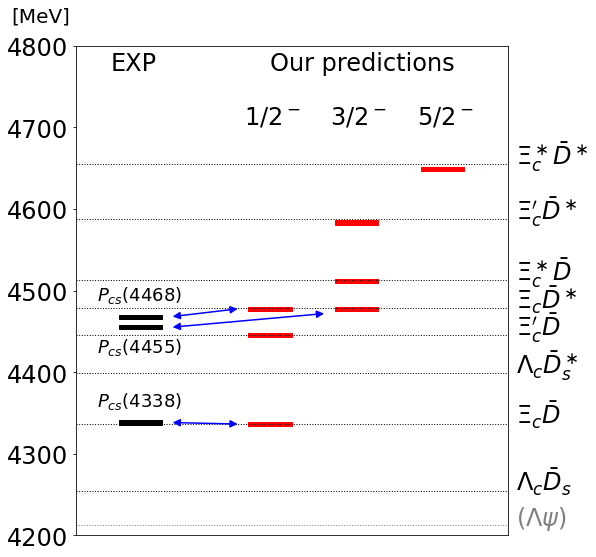}
\caption{
Comparison between experimental masses of $P_{cs}$ and theoretical predictions of our model when $F = 27$ is employed.  
The correspondence between the theoretical predictions and experimental data 
is denoted with arrows.  
}
\label{figresult}
\end{figure}

Having constructed the model, we show  the results for $P_{cs}$ 
in Fig.~\ref{figresult} in comparison with the existing experimental data.
The figure is made for $F = 27$ which forms a weakly bound state of $ \Xi_c \bar D$
with binding energy $E_B \sim 0.1$ MeV for $P_{cs}(4338)$.  
The very shallow bound state has a large spatial size $\sim 10$ fm, according to the formula 
for the root mean square radius of a weakly bound state, 
$\langle r^2\rangle^{1/2} = 1/(2\sqrt{\mu E_B})$ where $\mu$ is a reduced mass of the two-body system.  
Such a large state has only a small overlap with the interaction region $\sim 1$ fm$^3$  
affected by the $P_{cs}(4338) \to J/\psi \Lambda$ decay.  
This explains the small width of $P_{cs}(4338)$.  
Therefore, $P_{cs}(4338)$ is very likely to be a weakly bound molecular state
of $\Xi_c \bar D$ in an $S$-wave, with $J^P = 1/2^-$.

In Fig.~\ref{figresult}, in addition to that state, we find six more states with various spin and parity
$J^P = 1/2^-, 3/2^-, 5/2^-$.
All of these are molecular states of the near-threshold particles.  
By increasing the $F$ parameter, it is possible to lower the two predicted states located only slightly below the  
$\Xi_c \bar D^*$  threshold and a better agreement with  $P_{cs}(4455)$ and  $P_{cs}(4468)$ experimental masses is achieved.  
Numerical values of the obtained masses and widths for the above two cases are summarized in Table 2 in the supplementary material.
What is important is that our model predicts two states as $J^P = 1/2^-$ and $3/2^-$ molecules of  $\Xi_c(J=1/2)$ and  $\bar D^*(J=1)$ in the $S$-wave, supporting the two-peak interpretation of the experimental analysis by LHCb~\cite{Aaij:2020gdg}.
We suggest conducting a higher statistical data analysis in order to improve the statistical significance of those two states. 

The nature of these states that appear near threshold regions depends considerably on the attraction strength.
They may be either weakly bound or virtual states.  
Mathematically, the difference lies in the location of their poles; bound states are on the first Riemann sheet, while virtual states are on the second Riemann sheet.  
Whichever the case, the production rates of these states are amplified 
near the thresholds; from the experimental point of view this near-threshold amplification is a  physically important feature.

In addition to the above comparison with data, the present model contains important 
physics.
(1) The coupling to the compact five-quark components 
is effectively expressed as a short-range  attraction in the hadronic molecules. 
It is noticeable that such an interaction plays a dominant role in generating bound states. 
(2) The tensor force of the pion exchange causes $SD$-wave channel-couplings, which provides additional attraction.  
More interestingly, it controls decay widths, the inverse of the life time.
Without the tensor force, the decay width of, for instance, $\Xi_c^\prime \bar D^*$ ($3/2^-$) and $\Xi_c^\ast \bar{D}^\ast$ ($5/2^-$) molecules becomes smaller by one order of magnitude.

In hadronic systems, the above features are characteristic of those containing both heavy and light quarks, 
and hence are a result of the cooperation of chiral and heavy quark symmetries with colorful and colorless
forces of the strong interaction, QCD.  
These conditions have confirmed the conjecture regarding   the rich structure of 
hadronic molecules near the threshold, which was made almost half century 
ago~\cite{Gell-Mann:1964ewy,Zweig:1964ruk,DeRujula:1976ugc}.

The molecular structure near threshold region is a universal phenomenon of quantum systems 
that may appear in various matter hierarchies; quarks, hadrons (nuclei), atoms and molecules.  
Therefore, we expect to see interdisciplinary opportunities for various research activities to implement and discuss.

\section{Supplementary material}

The coupled-channel Hamiltonian for 
meson-baryon and five-quark channels is written in the form of block matrix as~\cite{Yamaguchi:2017zmn,Yamaguchi:2019seo}
\be
H = 
\begin{pmatrix}
H^{MB} & V \\
V^\dagger & H^{5q}
\end{pmatrix}
\ee
where $H^{MB}$ stands for meson-baryon ($MB$) channels, 
$H^{5q}$ for five-quark ($5q$) channels, and 
$V, V^\dagger$ their couplings.  
These are matrices whose dimensions are fixed by the number of 
base states (channels) of the meson-baryon and five-quark states.  
Explicitly, they are
\be
H^{MB}_{ij} 
&=&
\begin{pmatrix}
K_1 + V^m_{11} & V^m_{12} & \cdots \\
V^m_{21} & K_2 + V^m_{22} & \cdots \\
\cdots & \cdots & \cdots \
\end{pmatrix}
\nonumber \\
H^{5q}_{\alpha \beta} 
&=&
\begin{pmatrix}
M_1 & 0 & \cdots \\
0 & M_2  & \cdots \\
\cdots & \cdots & \cdots 
\end{pmatrix}
\ee
and 
\be
V_{i\alpha} = f \langle i | \alpha \rangle
=
\begin{pmatrix}
V_{11} & V_{12} & \cdots \\
V_{21} & V_{22}  & \cdots \\
\cdots & \cdots & \cdots 
\end{pmatrix}
\, .
\ee
In these equations, the label $m$ indicates the kind of mesons (either pion or kaon) 
exchanged between a meson and a baryon, $K_i$ the kinetic energy 
of the $i$-th  meson-baryon pair and $M_\alpha$ the masses of 
the $\alpha$-th five-quark channel.  
The couplings  of the meson-baryon and five-quark channels $V_{i\alpha}$ 
are expressed by the products of the overlap $\langle i | \alpha \rangle$
(spectroscopic factor) and the overall strength $f$.  
The overlap is computed when 
the channel's meson and baryon are 
in a region of interaction with  the five-quark state.  
This is a good working hypothesis, known in the study of cluster dynamics.  
The spectroscopic factor is obtained as the overlap of the color-spin-flavor wave functions of the meson-baryon and five-quark states, $\langle i|\alpha \rangle = \langle \phi^{i}_{MB}(CSF)|\phi^{\alpha}_{5q}(CSF) \rangle$, as discussed in Ref.~\cite{Yamaguchi:2017zmn}.

Setting the full-component wave function as 
$\psi = (\psi^{MB}, \psi^{5q})$, we obtain the coupled-channel equation, 
\be
H^{MB} \psi^{MB} + V\psi^{5q} &=& E \psi^{MB}
\, ,
\nonumber \\
V^\dagger 
 \psi^{MB} 
+ H^{5q}\psi^{5q} &=& E \psi^{5q}
\, .
\ee
By eliminating the five-quark channels (Feshbach's method~\cite{Feshbach:1958nx,Feshbach:1962ut}), 
we find 
the equation for the meson-baryon channels
\be
& & \left( K^{MB} + U \right)\psi^{MB}
= E \psi^{MB}, 
\nonumber \\
& & \ \ \ U = V^m + V \frac{1}{E-H^{5q}} V^\dagger 
\, .
\ee

The second term of the effective potential $U$ provides a short-range interaction that  is induced by a mixture of hadronic molecules and compact five-quark states. 
This effective potential consists of the meson-baryon One Meson Exchange Potential  (OMEP),  $V^m\equiv V^m_{ij}$, with $m=\pi$ or $K$ meson, and the coupling between the five-quark core configurations, $\alpha$, and the meson-baryon channels, $i$ , $ V\equiv V_{i\alpha}$. The explicit expressions of the OMEP and the  meson-baryon coupling values  are reported in Appendix~\ref{sec:appendix_OMEP}.
In general,  the effective potential $U$ is given as a non-local form with an energy dependence.
As discussed in Refs.~\cite{Yamaguchi:2017zmn,Yamaguchi:2019seo}, 
we reduce the complicated term to an energy-independent contact potential approximately.
This reduction is reasonable, if masses of the compact five-quark states are sufficiently larger than the threshold energies in which we are interested.

\subsection{Methods and numerical results}

We can solve 
the coupled-channel Schr\"odinger equation for meson-baryon 
states
by means of the Gaussian expansion method~\cite{Hiyama:2003cu} with the complex scaled coordinates~\cite{10.1143/PTP.116.1}, 
thereby finding their poles on the complex energy plane for the masses and decay widths 
for the pentaquarks.  

Expecting the lower partial-wave dominance for states near thresholds, 
we consider $S$-waves and $D,G$-waves that are coupled by the tensor force 
of the meson-exchange force.  

The numbers of channels depend on the quantum numbers, and are summarized in 
Table~\ref{table:coupled-channel_Pcs}.

Masses and decay widths of the pentaquarks predicted in this study are summarized in Table~\ref{table:Pcs_mass_width_B} for two coupling strengths, $F=27$ and $51$.
The former corresponds to Fig. 2 in the main text, while the latter, with the larger strength, shows the rich structure of hadronic molecules of $P_{cs}$.

\begin{table*}[htbp]
 \caption{\label{table:coupled-channel_Pcs} 
 Meson-baryon channels coupled to the hidden-charm strange pentaquarks $P_{cs}$ 
 of $J^P$ with $I=0$.
 The symbol $^{2S+1}L$ in the parentheses indicates possible spin ($S$) and orbital angular momentum $(L)$ of each meson-baryon channels.
 }
 \begin{center}
  \begin{tabular}{c|l}
   \toprule[0.3mm]
   $J^P$&Channels \\ 
   \midrule[0.3mm]
   $1/2^-$
   & $\Lambda_c\bar{D}_s(^2S),
       \Xi_c\bar{D}(^2S),
       \Lambda_c\bar{D}_s^\ast(^2S,^4D),
       \Xi_c^\prime\bar{D}(^2S),
       \Xi_c\bar{D}^\ast(^2S,^4D),
       \Xi_c^\ast\bar{D} (^4D),$ \\ 
   &$\Xi_c^\prime\bar{D}^\ast (^2S,^4D),
       \Xi_c^\ast\bar{D}^\ast(^2S,^4D,^6D)$ \\ \hline
  $3/2^-$
   &$\Lambda_c\bar{D}_s(^2D),
       \Xi_c\bar{D}(^2D),
       \Lambda_c\bar{D}_s^\ast(^4S,^2D,^4D),
       \Xi^\prime_c\bar{D}(^2D),
       \Xi_c\bar{D}^\ast(^4S,^2D,^4D),
       \Xi_c^\ast\bar{D}(^4S,^4D),$\\
    &$\Xi^\prime_c\bar{D}^\ast(^4S,^2D,^4D),
       \Xi_c^\ast\bar{D}^\ast(^4S,^2D,^4D,^6D,^6G)$ \\ 
   \hline
   $5/2^-$
   &$\Lambda_c\bar{D}_s(^2D),
       \Xi_c\bar{D}(^2D),
       \Lambda_c\bar{D}_s^\ast(^2D,^4D,^4G),
       \Xi^\prime_c\bar{D}(^2D),
       \Xi_c\bar{D}^\ast(^2D,^4D,^4G),
       \Xi_c^\ast\bar{D}(^4D,^4G), $ \\
   &$\Xi^\prime_c\bar{D}^\ast(^2D,^4D,^4G),
       \Xi_c^\ast\bar{D}^\ast(^6S,^2D,^4D,^6D,^4G,^6G)$ \\
   \bottomrule[0.3mm]
  \end{tabular}
 \end{center}
\end{table*}

\begin{table*}[htbp]
  \caption{\label{table:Pcs_mass_width_B}
 Comparison between the experimental masses and decay widths with our numerical results for 
 isospin $I=0$ in units of MeV.
 }
 \begin{center} 
  \begin{tabular}{c|ccc|ccc|ccc}
   \toprule[0.3mm]
   & & EXP~\cite{LHCbpentas2,LHCb:2020jpq}& & \multicolumn{3}{c|}{Our results for $F=27$} &
   \multicolumn{3}{c}{Our results for $F=51$}
   \\
   Threshold& State & Mass& Width& $J^P$ & Mass& Width 
			   & $J^P$ & Mass & Width\\
   \midrule[0.3mm]
   $\Lambda_c\bar{D}_s$&  
      ---&
	   ---
	   & 
	     ---
	       & 
		   --- &
		       --- &
			   --- &
			      $1/2^-$ &
   $4252.65$   & 
			 ---  \\ \hline
   $\Xi_c\bar{D}$& 
      $P_{cs}(4338)$ &
	   $4338.2$ 
	   & 
	     $7.0$  
	       & 
		   $1/2^-$ &
		       4336.34 &
			   $7.20 \times 10^{-2}$  &
			 $1/2^-$      &
   	   $4329.11$    
   &
	  1.54     
   \\ \hline
   $\Lambda_c \bar{D}_s^\ast$& 
      --- & 
	  --- & 
	      ---  &
		   --- &
		      ---
		       &
			   --- &
			  $1/2^-$ &
			       $4394.97$   & 
   $7.31\times 10^{-4}$
   \\ 
   & 
      --- & 
	  --- & 
	      ---  &
		   --- &
		       ---		       &			  
			   --- &
			  $3/2^-$ &
	    $4395.76$   
			        & 
    $8.78\times 10^{-4}$  
   \\ 
   \hline
   $\Xi_c^\prime \bar{D}$& 
      --- & 
	  --- & 
	      ---  &
		   $1/2^-$ &
		       4445.21  
		       &
			   0.341  
			   &
			  $1/2^-$ &
	    $4436.24$  
			       &
  	   2.12 
   \\ 
   \hline
   $\Xi_c\bar{D}^\ast$& 
      $P_{cs}(4455)$ & 
	  $4454.9$    
	   & 
	      $7.5$ 
	       & 
		   $3/2^-$ &
		       4476.92  
		       &
			   0.559  
			   &
			 $3/2^-$  &
			       $4465.24$   & 
   1.08   
   \\ 
   & 
      $P_{cs}(4468)$ &
	 $4467.8$  
	   & 
	     $5.2$ 
	       & 
		   $1/2^-$ &
		       4477.81 
		       &
			   0.210   
			   &
			  $1/2^-$ &
			       $4469.24$   & 
   2.31   
			   \\  \hline
   $\Xi_c^\ast\bar{D}$& 
      --- &
	 ---  & 
	    ---   & 
		   $3/2^-$ &
		       4511.98  
		       &
			   1.74   
			   &
			  $3/2^-$ &
			       $4502.91$   & 
	     4.09  
			   \\ \hline
   $\Xi_c^\prime\bar{D}^\ast$& 
      --- & 
	  ---  &
	      ---  & 
		   $3/2^-$ &
		       4583.29  
		       &
			   8.30  
			   &
			  $3/2^-$ &
			       $4567.12$  & 
   9.95   
			   \\ 
   $\Xi_c^\prime\bar{D}^\ast$& 
      --- & 
	  ---  &
	      ---  & 
		   ---&
		      --- &		
			   --- &
			  $1/2^-$ &
	  $4587.53$    & 
	   1.25    
			   \\ 
\hline
   $\Xi_c^\ast\bar{D}^\ast$& 
       --- & 
	   ---  & 
	       ---  &
		   $5/2^-$ &
		       4649.04 
		       &
			   12.4  
			   &
			  $5/2^-$ &
	  $4629.81$    & 
   14.7   
			   \\ 
   & 
       --- & 
	   ---  & 
	       ---  &
		   --- &
		       ---
		       &
			   --- &
			   $3/2^-$ &
			       $4653.02$   & 
   5.52  
			   \\ 
   \bottomrule[0.3mm]
  \end{tabular}
 \end{center}
\end{table*}

  \appendix


\subsection{Explicit form of the meson-baryon potential and couplings}
  \label{sec:appendix_OMEP}

  The OMEP for pion and kaon exchange used in this work are  the following:
\begin{widetext}
\begin{align}  
 V^\pi_{\bar{D}^{\ast} \Xi_c^\prime-\bar{D}\Xi_c} & =
  \frac{3 g_{\pi}g_1}{16\sqrt{3}f_\pi^2}\left[
 \vec{\varepsilon}\,^\dagger \cdot \vec{\sigma} C(r,m_\pi)
 + S_{\varepsilon \sigma}T(r,m_\pi) 
 \right] 
 , \\
   V^\pi_{\bar{D}^{\ast} \Xi_c^\ast-\bar{D}\Xi_c} & =
  -\frac{3g_{\pi}g_1}{16f_\pi^2}\left[
 \vec{\varepsilon}\,^\dagger \cdot \vec{\bar{\Sigma}} C(r,m_\pi)
 + S_{\varepsilon \bar{\Sigma}}T(r,m_\pi) 
 \right]
, \\
V^\pi_{\bar{D} \Xi_c^\prime-\bar{D}^\ast \Xi_c} & =
 \frac{3g_{\pi}g_1}{16\sqrt{3}f_\pi^2}\left[
 \vec{\varepsilon} \cdot \vec{\sigma} C(r,m_\pi)
 + S_{\varepsilon \sigma}T(r,m_\pi) 
 \right] 
, \\
V^\pi_{\bar{D} \Xi_c^\ast-\bar{D}^\ast \Xi_c} & =
 -\frac{3g_{\pi}g_1}{16f_\pi^2}\left[
 \vec{\varepsilon} \cdot \vec{\bar{\Sigma}} C(r,m_\pi)
 + S_{\varepsilon \bar{\Sigma}}T(r,m_\pi) 
 \right]  
, \\
V^\pi_{\bar{D}^\ast \Xi_c^\prime-\bar{D}^\ast \Xi_c} & =
  -\frac{3g_{\pi}g_1}{16\sqrt{3}f_\pi^2}\left[
 \vec{S} \cdot \vec{\sigma} C(r,m_\pi)
 + S_{S \sigma}T(r,m_\pi) 
 \right]  
 , \\
V^\pi_{\bar{D}^\ast \Xi_c^\ast-\bar{D}^\ast \Xi_c} & =
 \frac{3g_{\pi}g_1}{16f_\pi^2}\left[
 \vec{S} \cdot \vec{\bar{\Sigma}}\,^\dagger C(r,m_\pi)
 + S_{S \bar{\Sigma}}T(r,m_\pi) 
 \right]  
, \\
 V^\pi_{\bar{D}^{\ast} \Xi_c^\prime-\bar{D}\Xi_c^\prime} & =
\frac{g_{\pi}g_1}{8f_\pi^2}\left[
 \vec{\varepsilon}\,^\dagger \cdot \vec{\sigma} C(r,m_\pi)
 + S_{\varepsilon \sigma}T(r,m_\pi) 
 \right]  
 , \\
V^\pi_{\bar{D}^{\ast} \Xi_c^{\ast}-\bar{D}\Xi_c^\prime} & =
\frac{3g_{\pi}g_1}{16\sqrt{3}f_\pi^2}\left[
 \vec{\varepsilon}\,^\dagger \cdot \vec{\bar{\Sigma}}\,^\dagger C(r,m_\pi)
 + S_{\varepsilon \bar{\Sigma}}T(r,m_\pi) 
 \right] 
 , \\
V^\pi_{\bar{D}^{\ast} \Xi_c^\prime-\bar{D}\Xi_c^\ast} & =
 \frac{3g_{\pi}g_1}{16\sqrt{3}f_\pi^2}\left[
 \vec{\varepsilon}\,^\dagger \cdot \vec{\bar{\Sigma}} C(r,m_\pi)
 + S_{\varepsilon \bar{\Sigma}}T(r,m_\pi) 
 \right]  
 , \\
V^\pi_{\bar{D}^{\ast} \Xi_c^{\ast}-\bar{D}\Xi_c^\ast} & =
\frac{g_{\pi}g_1}{8f_\pi^2}\left[
 \vec{\varepsilon}\,^\dagger \cdot \vec{\Sigma} C(r,m_\pi)
 + S_{\varepsilon \Sigma}T(r,m_\pi) 
 \right]  
 , \\
   V^\pi_{\bar{D}^{\ast} \Xi_c^\prime-\bar{D}^\ast \Xi_c^\prime} & =
  -\frac{g_{\pi}g_1}{8f_\pi^2}\left[
 \vec{S} \cdot \vec{\sigma} C(r,m_\pi)
 + S_{S \sigma}T(r,m_\pi) 
 \right]  
 , \\
V^\pi_{\bar{D}^{\ast} \Xi_c^\ast-\bar{D}^\ast \Xi_c^\prime} & =
 -\frac{3g_{\pi}g_1}{16\sqrt{3}f_\pi^2}\left[
 \vec{S} \cdot \vec{\bar{\Sigma}}\,^\dagger C(r,m_\pi)
 + S_{S \bar{\Sigma}}T(r,m_\pi) 
 \right] 
 , \\
V^\pi_{\bar{D}^{\ast} \Xi_c^{\ast}-\bar{D}^\ast \Xi_c^\ast} & =
 -\frac{g_{\pi}g_1}{8f_\pi^2}\left[
 \vec{S} \cdot \vec{\Sigma} C(r,m_\pi)
 + S_{S \Sigma}T(r,m_\pi) 
 \right] 
 \, , \\
 V_{\bar{D}^{\ast} \Xi_c^\prime-\bar{D}_s\Lambda_c}^{K} & =
 -\frac{g_{\pi}g_1}{4\sqrt{6}f_\pi^2}\left[
 \vec{\varepsilon}\,^\dagger \cdot \vec{\sigma} C(r,m_K)
 + S_{\varepsilon \sigma}T(r,m_K) 
 \right]
 , \\
 V_{\bar{D}^{\ast} \Xi_c^\ast-\bar{D}_s\Lambda_c}^{K} & =
 \frac{g_{\pi}g_1}{4\sqrt{2}f_\pi^2}\left[
 \vec{\varepsilon}\,^\dagger \cdot \vec{\bar{\Sigma}} C(r,m_K)
 + S_{\varepsilon \bar{\Sigma}}T(r,m_K) 
 \right] 
, \\
 V_{\bar{D} \Xi_c^\prime-\bar{D}^\ast_s \Lambda_c}^{K} & =
 -\frac{g_{\pi}g_1}{4\sqrt{6}f_\pi^2}\left[
 \vec{\varepsilon} \cdot \vec{\sigma} C(r,m_K)
 + S_{\varepsilon \sigma}T(r,m_K) 
 \right]
, \\
V_{\bar{D} \Xi_c^\ast-\bar{D}^\ast_s \Lambda_c}^{K} & =
 \frac{g_{\pi}g_1}{4\sqrt{2}f_\pi^2}\left[
 \vec{\varepsilon} \cdot \vec{\bar{\Sigma}} C(r,m_K)
 + S_{\varepsilon \bar{\Sigma}}T(r,m_K) 
 \right] 
, \\
V_{\bar{D}^\ast \Xi_c^\prime-\bar{D}^\ast_s \Lambda_c}^{K} & =
 \frac{g_{\pi}g_1}{4\sqrt{6}f_\pi^2}\left[
 \vec{S} \cdot \vec{\sigma} C(r,m_K)
 + S_{S \sigma}T(r,m_K) 
 \right] 
 , \\
V_{\bar{D}^\ast \Xi_c^\ast-\bar{D}^\ast_s \Lambda_c}^{K} & =
 -\frac{g_{\pi}g_1}{4\sqrt{2}f_\pi^2}\left[
 \vec{S} \cdot \vec{\bar{\Sigma}}\,^\dagger C(r,m_K)
 + S_{S \bar{\Sigma}}T(r,m_K) 
 \right] 
 \, . 
\end{align}
\end{widetext}
$f_\pi$ is the pion decay constant given by $f_\pi=92.3$ MeV.
The coupling constant $g_\pi=0.59$ is determined by the strong decay of $D^\ast\to D\pi$~\cite{Manohar:2000dt}.
$g_1=1$ is estimated by the quark model~\cite{Liu:2011xc}.
$\vec{\varepsilon}$ and $\vec{\sigma}$ are the polarization vector and the Pauli matrices, respectively. The spin matrices $\vec{\bar{\Sigma}}$ are given by~\cite{Yamaguchi:2017zmn}
\begin{align}
 \bar{\Sigma}^\mu = \left(
 \begin{array}{cccc}
   \vec{\varepsilon}\,^{(+)} &\sqrt{2/3} \vec{\varepsilon}\,^{(0)} &\sqrt{1/3} \vec{\varepsilon}\,^{(-)} &0 \\
  0& \sqrt{1/3} \vec{\varepsilon}\,^{(+)}&\sqrt{2/3} \vec{\varepsilon}\,^{(0)} & \vec{\varepsilon}\,^{(-)} \\
 \end{array} 
 \right)^\mu \, .
\end{align}
$\vec{S}$ and $\vec{\Sigma}$ are obtained by
$\vec{S}=i\vec{\varepsilon}\times \vec{\varepsilon}\,^\dagger$ 
and 
$\vec{\Sigma}=(3/2) i \vec{\bar{\Sigma}}\times \vec{\bar{\Sigma}}\,^\dagger$, respectively.
The tensor operator $S_{{\cal O}_1{\cal O}_2}(\hat{r})$ is defined by $S_{{\cal O}_1{\cal O}_2}(\hat{r})=3 \vec{\cal O}_1\cdot\hat{r} \vec{\cal O}_2\cdot\hat{r} - \vec{\cal O}_1\cdot\vec{\cal O}_2$.
The functions $C(r,m)$ and $T(r,m)$ are defined by
\begin{align}
  C(r,m) = & \int \frac{d^3 q}{(2\pi)^3} \frac{m^2}{\vec{q}\,^2+m^2} e^{i\vec{q}\cdot\vec{r}} 
 \notag\\
 & \times F_M(\Lambda, m, \vec{q}\,)  F_B(\Lambda, m, \vec{q}\,) 
 \, , \\
  S_{{\cal O}_1{\cal O}_2}(\hat{r})T(r,m) = & 
 \int \frac{d^3 q}{(2\pi)^3} \frac{-\vec{q}\,^2}{\vec{q}\,^2+m^2} S_{{\cal O}_1{\cal O}_2}(\hat{q}) e^{i\vec{q}\cdot\vec{r}}
 \notag\\
 & \times F_M(\Lambda, m,\vec{q}\,)  F_B(\Lambda, m,\vec{q}\,)
 \, , 
\end{align}
where $F_i(\Lambda, m, \vec{q}\,)$ ($i=M,B$) is the form factor introduced at each vertex:
\begin{align}
 F_i(\Lambda, m, \vec{q}\,) = 
 \frac{\Lambda_i^2-m^2}{\Lambda_i^2+\vec{q}\,^2} \,  .
\end{align}
As discussed in~\cite{Yamaguchi:2017zmn},
the cutoff parameters $\Lambda_i$ are determined by the size ratio between the heavy meson and nucleon, $\Lambda_N/\Lambda_i = r_i/r_N$.
With the nucleon cutoff $\Lambda_N=837$ MeV,  we obtain $\Lambda_M=1.35\Lambda_N$ for the heavy mesons, and $\Lambda_B=\Lambda_N$ for the heavy baryons.

\bibliography{reference}

\begin{thebibliography}{36}%
\makeatletter
\providecommand \@ifxundefined [1]{%
 \@ifx{#1\undefined}
}%
\providecommand \@ifnum [1]{%
 \ifnum #1\expandafter \@firstoftwo
 \else \expandafter \@secondoftwo
 \fi
}%
\providecommand \@ifx [1]{%
 \ifx #1\expandafter \@firstoftwo
 \else \expandafter \@secondoftwo
 \fi
}%
\providecommand \natexlab [1]{#1}%
\providecommand \enquote  [1]{``#1''}%
\providecommand \bibnamefont  [1]{#1}%
\providecommand \bibfnamefont [1]{#1}%
\providecommand \citenamefont [1]{#1}%
\providecommand \href@noop [0]{\@secondoftwo}%
\providecommand \href [0]{\begingroup \@sanitize@url \@href}%
\providecommand \@href[1]{\@@startlink{#1}\@@href}%
\providecommand \@@href[1]{\endgroup#1\@@endlink}%
\providecommand \@sanitize@url [0]{\catcode `\\12\catcode `\$12\catcode
  `\&12\catcode `\#12\catcode `\^12\catcode `\_12\catcode `\%12\relax}%
\providecommand \@@startlink[1]{}%
\providecommand \@@endlink[0]{}%
\providecommand \url  [0]{\begingroup\@sanitize@url \@url }%
\providecommand \@url [1]{\endgroup\@href {#1}{\urlprefix }}%
\providecommand \urlprefix  [0]{URL }%
\providecommand \Eprint [0]{\href }%
\providecommand \doibase [0]{http://dx.doi.org/}%
\providecommand \selectlanguage [0]{\@gobble}%
\providecommand \bibinfo  [0]{\@secondoftwo}%
\providecommand \bibfield  [0]{\@secondoftwo}%
\providecommand \translation [1]{[#1]}%
\providecommand \BibitemOpen [0]{}%
\providecommand \bibitemStop [0]{}%
\providecommand \bibitemNoStop [0]{.\EOS\space}%
\providecommand \EOS [0]{\spacefactor3000\relax}%
\providecommand \BibitemShut  [1]{\csname bibitem#1\endcsname}%
\let\auto@bib@innerbib\@empty
\bibitem [{\citenamefont {Gell-Mann}(1964)}]{Gell-Mann:1964ewy}%
  \BibitemOpen
  \bibfield  {author} {\bibinfo {author} {\bibfnamefont {M.}~\bibnamefont
  {Gell-Mann}},\ }\href {\doibase 10.1016/S0031-9163(64)92001-3} {\bibfield
  {journal} {\bibinfo  {journal} {Phys. Lett.}\ }\textbf {\bibinfo {volume}
  {8}},\ \bibinfo {pages} {214} (\bibinfo {year} {1964})}\BibitemShut {NoStop}%
\bibitem [{\citenamefont {Zweig}(1964)}]{Zweig:1964ruk}%
  \BibitemOpen
  \bibfield  {author} {\bibinfo {author} {\bibfnamefont {G.}~\bibnamefont
  {Zweig}},\ }\bibfield  {booktitle} {\emph {\bibinfo {booktitle} {{An SU(3)
  model for strong interaction symmetry and its breaking. Version 1}}},\
  }\href@noop {} {\bibfield  {journal} {\bibinfo  {journal} {CERN-TH-401}\ }
  (\bibinfo {year} {1964})}\BibitemShut {NoStop}%
\bibitem [{\citenamefont {De~Rujula}\ \emph {et~al.}(1976)\citenamefont
  {De~Rujula}, \citenamefont {Georgi},\ and\ \citenamefont
  {Glashow}}]{DeRujula:1976ugc}%
  \BibitemOpen
  \bibfield  {author} {\bibinfo {author} {\bibfnamefont {A.}~\bibnamefont
  {De~Rujula}}, \bibinfo {author} {\bibfnamefont {H.}~\bibnamefont {Georgi}}, \
  and\ \bibinfo {author} {\bibfnamefont {S.~L.}\ \bibnamefont {Glashow}},\
  }\href {\doibase 10.1103/PhysRevLett.37.785} {\bibfield  {journal} {\bibinfo
  {journal} {Phys. Rev. Lett.}\ }\textbf {\bibinfo {volume} {37}},\ \bibinfo
  {pages} {785} (\bibinfo {year} {1976})}\BibitemShut {NoStop}%
\bibitem [{\citenamefont {Choi}\ \emph {et~al.}(2003)\citenamefont {Choi} \emph
  {et~al.}}]{Belle:2003nnu}%
  \BibitemOpen
  \bibfield  {author} {\bibinfo {author} {\bibfnamefont {S.~K.}\ \bibnamefont
  {Choi}} \emph {et~al.} (\bibinfo {collaboration} {Belle}),\ }\href {\doibase
  10.1103/PhysRevLett.91.262001} {\bibfield  {journal} {\bibinfo  {journal}
  {Phys. Rev. Lett.}\ }\textbf {\bibinfo {volume} {91}},\ \bibinfo {pages}
  {262001} (\bibinfo {year} {2003})},\ \Eprint
  {http://arxiv.org/abs/hep-ex/0309032} {arXiv:hep-ex/0309032} \BibitemShut
  {NoStop}%
\bibitem [{\citenamefont {Spadaro~Norella}\ and\ \citenamefont
  {Chen}()}]{LHCbpentas2}%
  \BibitemOpen
  \bibfield  {author} {\bibinfo {author} {\bibfnamefont {E.}~\bibnamefont
  {Spadaro~Norella}}\ and\ \bibinfo {author} {\bibfnamefont {C.}~\bibnamefont
  {Chen}} (\bibinfo {collaboration} {LHCb}),\ }\bibfield  {booktitle} {\emph
  {\bibinfo {booktitle} {{Particle Zoo 2.0: New Tetra- and Pentaquarks at
  LHCb}}},\ }\href@noop {} {\bibinfo  {journal} {CERN Seminar, 5th July
  (2022)}\ }\BibitemShut {NoStop}%
\bibitem [{\citenamefont {Brambilla}\ \emph {et~al.}(2022)\citenamefont
  {Brambilla} \emph {et~al.}}]{Brambilla:2022ura}%
  \BibitemOpen
\bibfield  {journal} {  }\bibfield  {author} {\bibinfo {author} {\bibfnamefont
  {N.}~\bibnamefont {Brambilla}} \emph {et~al.},\ }\href@noop {} {\  (\bibinfo
  {year} {2022})},\ \Eprint {http://arxiv.org/abs/2203.16583} {arXiv:2203.16583
  [hep-ph]} \BibitemShut {NoStop}%
\bibitem [{\citenamefont {Aaij}\ \emph {et~al.}(2015)\citenamefont {Aaij} \emph
  {et~al.}}]{Aaij:2015tga}%
  \BibitemOpen
  \bibfield  {author} {\bibinfo {author} {\bibfnamefont {R.}~\bibnamefont
  {Aaij}} \emph {et~al.} (\bibinfo {collaboration} {LHCb}),\ }\href {\doibase
  10.1103/PhysRevLett.115.072001} {\bibfield  {journal} {\bibinfo  {journal}
  {Phys. Rev. Lett.}\ }\textbf {\bibinfo {volume} {115}},\ \bibinfo {pages}
  {072001} (\bibinfo {year} {2015})},\ \Eprint
  {http://arxiv.org/abs/1507.03414} {arXiv:1507.03414 [hep-ex]} \BibitemShut
  {NoStop}%
\bibitem [{\citenamefont {Aaij}\ \emph {et~al.}(2019)\citenamefont {Aaij} \emph
  {et~al.}}]{Aaij:2019vzc}%
  \BibitemOpen
  \bibfield  {author} {\bibinfo {author} {\bibfnamefont {R.}~\bibnamefont
  {Aaij}} \emph {et~al.} (\bibinfo {collaboration} {LHCb}),\ }\href {\doibase
  10.1103/PhysRevLett.122.222001} {\bibfield  {journal} {\bibinfo  {journal}
  {Phys. Rev. Lett.}\ }\textbf {\bibinfo {volume} {122}},\ \bibinfo {pages}
  {222001} (\bibinfo {year} {2019})},\ \Eprint
  {http://arxiv.org/abs/1904.03947} {arXiv:1904.03947 [hep-ex]} \BibitemShut
  {NoStop}%
\bibitem [{\citenamefont {Aaij}\ \emph
  {et~al.}(2022{\natexlab{a}})\citenamefont {Aaij} \emph
  {et~al.}}]{LHCb:2021chn}%
  \BibitemOpen
  \bibfield  {author} {\bibinfo {author} {\bibfnamefont {R.}~\bibnamefont
  {Aaij}} \emph {et~al.} (\bibinfo {collaboration} {LHCb}),\ }\href {\doibase
  10.1103/PhysRevLett.128.062001} {\bibfield  {journal} {\bibinfo  {journal}
  {Phys. Rev. Lett.}\ }\textbf {\bibinfo {volume} {128}},\ \bibinfo {pages}
  {062001} (\bibinfo {year} {2022}{\natexlab{a}})},\ \Eprint
  {http://arxiv.org/abs/2108.04720} {arXiv:2108.04720 [hep-ex]} \BibitemShut
  {NoStop}%
\bibitem [{\citenamefont {Aaij}\ \emph
  {et~al.}(2021{\natexlab{a}})\citenamefont {Aaij} \emph
  {et~al.}}]{Aaij:2020gdg}%
  \BibitemOpen
  \bibfield  {author} {\bibinfo {author} {\bibfnamefont {R.}~\bibnamefont
  {Aaij}} \emph {et~al.} (\bibinfo {collaboration} {LHCb}),\ }\href {\doibase
  10.1016/j.scib.2021.02.030} {\bibfield  {journal} {\bibinfo  {journal} {Sci.
  Bull.}\ }\textbf {\bibinfo {volume} {66}},\ \bibinfo {pages} {1278} (\bibinfo
  {year} {2021}{\natexlab{a}})},\ \Eprint {http://arxiv.org/abs/2012.10380}
  {arXiv:2012.10380 [hep-ex]} \BibitemShut {NoStop}%
\bibitem [{\citenamefont {Karliner}\ and\ \citenamefont
  {Rosner}(2022)}]{Karliner:2022erb}%
  \BibitemOpen
  \bibfield  {author} {\bibinfo {author} {\bibfnamefont {M.}~\bibnamefont
  {Karliner}}\ and\ \bibinfo {author} {\bibfnamefont {J.~R.}\ \bibnamefont
  {Rosner}},\ }\href@noop {} {\  (\bibinfo {year} {2022})},\ \Eprint
  {http://arxiv.org/abs/2207.07581} {arXiv:2207.07581 [hep-ph]} \BibitemShut
  {NoStop}%
\bibitem [{\citenamefont {Aaij}\ \emph
  {et~al.}(2022{\natexlab{b}})\citenamefont {Aaij} \emph
  {et~al.}}]{LHCb:2021vvq}%
  \BibitemOpen
  \bibfield  {author} {\bibinfo {author} {\bibfnamefont {R.}~\bibnamefont
  {Aaij}} \emph {et~al.} (\bibinfo {collaboration} {LHCb}),\ }\href {\doibase
  10.1038/s41567-022-01614-y} {\bibfield  {journal} {\bibinfo  {journal}
  {Nature Phys.}\ }\textbf {\bibinfo {volume} {18}},\ \bibinfo {pages} {751}
  (\bibinfo {year} {2022}{\natexlab{b}})},\ \Eprint
  {http://arxiv.org/abs/2109.01038} {arXiv:2109.01038 [hep-ex]} \BibitemShut
  {NoStop}%
\bibitem [{\citenamefont {Aaij}\ \emph
  {et~al.}(2022{\natexlab{c}})\citenamefont {Aaij} \emph
  {et~al.}}]{LHCb:2021auc}%
  \BibitemOpen
  \bibfield  {author} {\bibinfo {author} {\bibfnamefont {R.}~\bibnamefont
  {Aaij}} \emph {et~al.} (\bibinfo {collaboration} {LHCb}),\ }\href {\doibase
  10.1038/s41467-022-30206-w} {\bibfield  {journal} {\bibinfo  {journal}
  {Nature Commun.}\ }\textbf {\bibinfo {volume} {13}},\ \bibinfo {pages} {3351}
  (\bibinfo {year} {2022}{\natexlab{c}})},\ \Eprint
  {http://arxiv.org/abs/2109.01056} {arXiv:2109.01056 [hep-ex]} \BibitemShut
  {NoStop}%
\bibitem [{\citenamefont {Xiao}\ \emph {et~al.}(2019)\citenamefont {Xiao},
  \citenamefont {Nieves},\ and\ \citenamefont {Oset}}]{Xiao:2019gjd}%
  \BibitemOpen
  \bibfield  {author} {\bibinfo {author} {\bibfnamefont {C.~W.}\ \bibnamefont
  {Xiao}}, \bibinfo {author} {\bibfnamefont {J.}~\bibnamefont {Nieves}}, \ and\
  \bibinfo {author} {\bibfnamefont {E.}~\bibnamefont {Oset}},\ }\href {\doibase
  10.1016/j.physletb.2019.135051} {\bibfield  {journal} {\bibinfo  {journal}
  {Phys. Lett. B}\ }\textbf {\bibinfo {volume} {799}},\ \bibinfo {pages}
  {135051} (\bibinfo {year} {2019})},\ \Eprint
  {http://arxiv.org/abs/1906.09010} {arXiv:1906.09010 [hep-ph]} \BibitemShut
  {NoStop}%
\bibitem [{\citenamefont {Wang}\ \emph {et~al.}(2020)\citenamefont {Wang},
  \citenamefont {Meng},\ and\ \citenamefont {Zhu}}]{Wang:2019nvm}%
  \BibitemOpen
  \bibfield  {author} {\bibinfo {author} {\bibfnamefont {B.}~\bibnamefont
  {Wang}}, \bibinfo {author} {\bibfnamefont {L.}~\bibnamefont {Meng}}, \ and\
  \bibinfo {author} {\bibfnamefont {S.-L.}\ \bibnamefont {Zhu}},\ }\href
  {\doibase 10.1103/PhysRevD.101.034018} {\bibfield  {journal} {\bibinfo
  {journal} {Phys. Rev. D}\ }\textbf {\bibinfo {volume} {101}},\ \bibinfo
  {pages} {034018} (\bibinfo {year} {2020})},\ \Eprint
  {http://arxiv.org/abs/1912.12592} {arXiv:1912.12592 [hep-ph]} \BibitemShut
  {NoStop}%
\bibitem [{\citenamefont {Chen}\ \emph {et~al.}(2021)\citenamefont {Chen},
  \citenamefont {Chen}, \citenamefont {Liu},\ and\ \citenamefont
  {Liu}}]{Chen:2020uif}%
  \BibitemOpen
  \bibfield  {author} {\bibinfo {author} {\bibfnamefont {H.-X.}\ \bibnamefont
  {Chen}}, \bibinfo {author} {\bibfnamefont {W.}~\bibnamefont {Chen}}, \bibinfo
  {author} {\bibfnamefont {X.}~\bibnamefont {Liu}}, \ and\ \bibinfo {author}
  {\bibfnamefont {X.-H.}\ \bibnamefont {Liu}},\ }\href {\doibase
  10.1140/epjc/s10052-021-09196-4} {\bibfield  {journal} {\bibinfo  {journal}
  {Eur. Phys. J. C}\ }\textbf {\bibinfo {volume} {81}},\ \bibinfo {pages} {409}
  (\bibinfo {year} {2021})},\ \Eprint {http://arxiv.org/abs/2011.01079}
  {arXiv:2011.01079 [hep-ph]} \BibitemShut {NoStop}%
\bibitem [{\citenamefont {Ali}\ \emph {et~al.}(2019)\citenamefont {Ali},
  \citenamefont {Ahmed}, \citenamefont {Aslam}, \citenamefont {Parkhomenko},\
  and\ \citenamefont {Rehman}}]{Ali:2019clg}%
  \BibitemOpen
  \bibfield  {author} {\bibinfo {author} {\bibfnamefont {A.}~\bibnamefont
  {Ali}}, \bibinfo {author} {\bibfnamefont {I.}~\bibnamefont {Ahmed}}, \bibinfo
  {author} {\bibfnamefont {M.~J.}\ \bibnamefont {Aslam}}, \bibinfo {author}
  {\bibfnamefont {A.~Y.}\ \bibnamefont {Parkhomenko}}, \ and\ \bibinfo {author}
  {\bibfnamefont {A.}~\bibnamefont {Rehman}},\ }\href {\doibase
  10.1007/JHEP10(2019)256} {\bibfield  {journal} {\bibinfo  {journal} {JHEP}\
  }\textbf {\bibinfo {volume} {10}},\ \bibinfo {pages} {256} (\bibinfo {year}
  {2019})},\ \Eprint {http://arxiv.org/abs/1907.06507} {arXiv:1907.06507
  [hep-ph]} \BibitemShut {NoStop}%
\bibitem [{\citenamefont {Ferretti}\ and\ \citenamefont
  {Santopinto}(2022)}]{Ferretti:2021zis}%
  \BibitemOpen
  \bibfield  {author} {\bibinfo {author} {\bibfnamefont {J.}~\bibnamefont
  {Ferretti}}\ and\ \bibinfo {author} {\bibfnamefont {E.}~\bibnamefont
  {Santopinto}},\ }\href {\doibase 10.1016/j.scib.2022.04.010} {\bibfield
  {journal} {\bibinfo  {journal} {Sci. Bull.}\ }\textbf {\bibinfo {volume}
  {67}},\ \bibinfo {pages} {1209} (\bibinfo {year} {2022})},\ \Eprint
  {http://arxiv.org/abs/2111.08650} {arXiv:2111.08650 [hep-ph]} \BibitemShut
  {NoStop}%
\bibitem [{\citenamefont {Chen}\ and\ \citenamefont
  {Liu}(2022)}]{Chen:2022onm}%
  \BibitemOpen
  \bibfield  {author} {\bibinfo {author} {\bibfnamefont {R.}~\bibnamefont
  {Chen}}\ and\ \bibinfo {author} {\bibfnamefont {X.}~\bibnamefont {Liu}},\
  }\href {\doibase 10.1103/PhysRevD.105.014029} {\bibfield  {journal} {\bibinfo
   {journal} {Phys. Rev. D}\ }\textbf {\bibinfo {volume} {105}},\ \bibinfo
  {pages} {014029} (\bibinfo {year} {2022})},\ \Eprint
  {http://arxiv.org/abs/2201.07603} {arXiv:2201.07603 [hep-ph]} \BibitemShut
  {NoStop}%
\bibitem [{\citenamefont {Wang}\ and\ \citenamefont
  {Liu}(2022)}]{Wang:2022mxy}%
  \BibitemOpen
  \bibfield  {author} {\bibinfo {author} {\bibfnamefont {F.-L.}\ \bibnamefont
  {Wang}}\ and\ \bibinfo {author} {\bibfnamefont {X.}~\bibnamefont {Liu}},\
  }\href@noop {} {\  (\bibinfo {year} {2022})},\ \Eprint
  {http://arxiv.org/abs/2207.10493} {arXiv:2207.10493 [hep-ph]} \BibitemShut
  {NoStop}%
\bibitem [{\citenamefont {Yan}\ \emph {et~al.}(2022)\citenamefont {Yan},
  \citenamefont {Peng}, \citenamefont {S\'anchez~S\'anchez},\ and\
  \citenamefont {Pavon~Valderrama}}]{Yan:2022wuz}%
  \BibitemOpen
  \bibfield  {author} {\bibinfo {author} {\bibfnamefont {M.-J.}\ \bibnamefont
  {Yan}}, \bibinfo {author} {\bibfnamefont {F.-Z.}\ \bibnamefont {Peng}},
  \bibinfo {author} {\bibfnamefont {M.}~\bibnamefont {S\'anchez~S\'anchez}}, \
  and\ \bibinfo {author} {\bibfnamefont {M.}~\bibnamefont {Pavon~Valderrama}},\
  }\href@noop {} {\  (\bibinfo {year} {2022})},\ \Eprint
  {http://arxiv.org/abs/2207.11144} {arXiv:2207.11144 [hep-ph]} \BibitemShut
  {NoStop}%
\bibitem [{\citenamefont {Burns}\ and\ \citenamefont
  {Swanson}(2022)}]{Burns:2022uha}%
  \BibitemOpen
  \bibfield  {author} {\bibinfo {author} {\bibfnamefont {T.~J.}\ \bibnamefont
  {Burns}}\ and\ \bibinfo {author} {\bibfnamefont {E.~S.}\ \bibnamefont
  {Swanson}},\ }\href@noop {} {\  (\bibinfo {year} {2022})},\ \Eprint
  {http://arxiv.org/abs/2208.05106} {arXiv:2208.05106 [hep-ph]} \BibitemShut
  {NoStop}%
\bibitem [{\citenamefont {Collaboration}\ \emph {et~al.}(2020)\citenamefont
  {Collaboration} \emph {et~al.}}]{ALICE:2020mfd}%
  \BibitemOpen
  \bibfield  {author} {\bibinfo {author} {\bibfnamefont {A.}~\bibnamefont
  {Collaboration}} \emph {et~al.} (\bibinfo {collaboration} {ALICE}),\ }\href
  {\doibase 10.1038/s41586-020-3001-6} {\bibfield  {journal} {\bibinfo
  {journal} {Nature}\ }\textbf {\bibinfo {volume} {588}},\ \bibinfo {pages}
  {232} (\bibinfo {year} {2020})},\ \bibinfo {note} {[Erratum: Nature 590, E13
  (2021)]},\ \Eprint {http://arxiv.org/abs/2005.11495} {arXiv:2005.11495
  [nucl-ex]} \BibitemShut {NoStop}%
\bibitem [{\citenamefont {Hatsuda}(2018)}]{Hatsuda:2018nes}%
  \BibitemOpen
  \bibfield  {author} {\bibinfo {author} {\bibfnamefont {T.}~\bibnamefont
  {Hatsuda}},\ }\href {\doibase 10.1007/s11467-018-0829-4} {\bibfield
  {journal} {\bibinfo  {journal} {Front. Phys. (Beijing)}\ }\textbf {\bibinfo
  {volume} {13}},\ \bibinfo {pages} {132105} (\bibinfo {year}
  {2018})}\BibitemShut {NoStop}%
\bibitem [{\citenamefont {Yamaguchi}\ \emph
  {et~al.}(2020{\natexlab{a}})\citenamefont {Yamaguchi}, \citenamefont
  {Hosaka}, \citenamefont {Takeuchi},\ and\ \citenamefont
  {Takizawa}}]{Yamaguchi:2019vea}%
  \BibitemOpen
  \bibfield  {author} {\bibinfo {author} {\bibfnamefont {Y.}~\bibnamefont
  {Yamaguchi}}, \bibinfo {author} {\bibfnamefont {A.}~\bibnamefont {Hosaka}},
  \bibinfo {author} {\bibfnamefont {S.}~\bibnamefont {Takeuchi}}, \ and\
  \bibinfo {author} {\bibfnamefont {M.}~\bibnamefont {Takizawa}},\ }\href
  {\doibase 10.1088/1361-6471/ab72b0} {\bibfield  {journal} {\bibinfo
  {journal} {J. Phys. G}\ }\textbf {\bibinfo {volume} {47}},\ \bibinfo {pages}
  {053001} (\bibinfo {year} {2020}{\natexlab{a}})},\ \Eprint
  {http://arxiv.org/abs/1908.08790} {arXiv:1908.08790 [hep-ph]} \BibitemShut
  {NoStop}%
\bibitem [{\citenamefont {Santopinto}\ and\ \citenamefont
  {Giachino}(2017)}]{Santopinto:2016pkp}%
  \BibitemOpen
  \bibfield  {author} {\bibinfo {author} {\bibfnamefont {E.}~\bibnamefont
  {Santopinto}}\ and\ \bibinfo {author} {\bibfnamefont {A.}~\bibnamefont
  {Giachino}},\ }\href {\doibase 10.1103/PhysRevD.96.014014} {\bibfield
  {journal} {\bibinfo  {journal} {Phys. Rev. D}\ }\textbf {\bibinfo {volume}
  {96}},\ \bibinfo {pages} {014014} (\bibinfo {year} {2017})},\ \Eprint
  {http://arxiv.org/abs/1604.03769} {arXiv:1604.03769 [hep-ph]} \BibitemShut
  {NoStop}%
\bibitem [{\citenamefont {Takeuchi}\ and\ \citenamefont
  {Takizawa}(2017)}]{Takeuchi:2016ejt}%
  \BibitemOpen
  \bibfield  {author} {\bibinfo {author} {\bibfnamefont {S.}~\bibnamefont
  {Takeuchi}}\ and\ \bibinfo {author} {\bibfnamefont {M.}~\bibnamefont
  {Takizawa}},\ }\href {\doibase 10.1016/j.physletb.2016.11.034} {\bibfield
  {journal} {\bibinfo  {journal} {Phys. Lett. B}\ }\textbf {\bibinfo {volume}
  {764}},\ \bibinfo {pages} {254} (\bibinfo {year} {2017})},\ \Eprint
  {http://arxiv.org/abs/1608.05475} {arXiv:1608.05475 [hep-ph]} \BibitemShut
  {NoStop}%
\bibitem [{\citenamefont {Yamaguchi}\ \emph {et~al.}(2017)\citenamefont
  {Yamaguchi}, \citenamefont {Giachino}, \citenamefont {Hosaka}, \citenamefont
  {Santopinto}, \citenamefont {Takeuchi},\ and\ \citenamefont
  {Takizawa}}]{Yamaguchi:2017zmn}%
  \BibitemOpen
  \bibfield  {author} {\bibinfo {author} {\bibfnamefont {Y.}~\bibnamefont
  {Yamaguchi}}, \bibinfo {author} {\bibfnamefont {A.}~\bibnamefont {Giachino}},
  \bibinfo {author} {\bibfnamefont {A.}~\bibnamefont {Hosaka}}, \bibinfo
  {author} {\bibfnamefont {E.}~\bibnamefont {Santopinto}}, \bibinfo {author}
  {\bibfnamefont {S.}~\bibnamefont {Takeuchi}}, \ and\ \bibinfo {author}
  {\bibfnamefont {M.}~\bibnamefont {Takizawa}},\ }\href {\doibase
  10.1103/PhysRevD.96.114031} {\bibfield  {journal} {\bibinfo  {journal} {Phys.
  Rev. D}\ }\textbf {\bibinfo {volume} {96}},\ \bibinfo {pages} {114031}
  (\bibinfo {year} {2017})},\ \Eprint {http://arxiv.org/abs/1709.00819}
  {arXiv:1709.00819 [hep-ph]} \BibitemShut {NoStop}%
\bibitem [{\citenamefont {Yamaguchi}\ \emph
  {et~al.}(2020{\natexlab{b}})\citenamefont {Yamaguchi}, \citenamefont
  {Garc\'\i{}a-Tecocoatzi}, \citenamefont {Giachino}, \citenamefont {Hosaka},
  \citenamefont {Santopinto}, \citenamefont {Takeuchi},\ and\ \citenamefont
  {Takizawa}}]{Yamaguchi:2019seo}%
  \BibitemOpen
  \bibfield  {author} {\bibinfo {author} {\bibfnamefont {Y.}~\bibnamefont
  {Yamaguchi}}, \bibinfo {author} {\bibfnamefont {H.}~\bibnamefont
  {Garc\'\i{}a-Tecocoatzi}}, \bibinfo {author} {\bibfnamefont {A.}~\bibnamefont
  {Giachino}}, \bibinfo {author} {\bibfnamefont {A.}~\bibnamefont {Hosaka}},
  \bibinfo {author} {\bibfnamefont {E.}~\bibnamefont {Santopinto}}, \bibinfo
  {author} {\bibfnamefont {S.}~\bibnamefont {Takeuchi}}, \ and\ \bibinfo
  {author} {\bibfnamefont {M.}~\bibnamefont {Takizawa}},\ }\href {\doibase
  10.1103/PhysRevD.101.091502} {\bibfield  {journal} {\bibinfo  {journal}
  {Phys. Rev. D}\ }\textbf {\bibinfo {volume} {101}},\ \bibinfo {pages}
  {091502} (\bibinfo {year} {2020}{\natexlab{b}})},\ \Eprint
  {http://arxiv.org/abs/1907.04684} {arXiv:1907.04684 [hep-ph]} \BibitemShut
  {NoStop}%
\bibitem [{\citenamefont {Feshbach}(1958)}]{Feshbach:1958nx}%
  \BibitemOpen
  \bibfield  {author} {\bibinfo {author} {\bibfnamefont {H.}~\bibnamefont
  {Feshbach}},\ }\href {\doibase 10.1016/0003-4916(58)90007-1} {\bibfield
  {journal} {\bibinfo  {journal} {Annals Phys.}\ }\textbf {\bibinfo {volume}
  {5}},\ \bibinfo {pages} {357} (\bibinfo {year} {1958})}\BibitemShut {NoStop}%
\bibitem [{\citenamefont {Feshbach}(1962)}]{Feshbach:1962ut}%
  \BibitemOpen
  \bibfield  {author} {\bibinfo {author} {\bibfnamefont {H.}~\bibnamefont
  {Feshbach}},\ }\href {\doibase 10.1016/0003-4916(62)90221-X} {\bibfield
  {journal} {\bibinfo  {journal} {Annals Phys.}\ }\textbf {\bibinfo {volume}
  {19}},\ \bibinfo {pages} {287} (\bibinfo {year} {1962})}\BibitemShut
  {NoStop}%
\bibitem [{\citenamefont {Hiyama}\ \emph {et~al.}(2003)\citenamefont {Hiyama},
  \citenamefont {Kino},\ and\ \citenamefont {Kamimura}}]{Hiyama:2003cu}%
  \BibitemOpen
  \bibfield  {author} {\bibinfo {author} {\bibfnamefont {E.}~\bibnamefont
  {Hiyama}}, \bibinfo {author} {\bibfnamefont {Y.}~\bibnamefont {Kino}}, \ and\
  \bibinfo {author} {\bibfnamefont {M.}~\bibnamefont {Kamimura}},\ }\href
  {\doibase 10.1016/S0146-6410(03)90015-9} {\bibfield  {journal} {\bibinfo
  {journal} {Prog. Part. Nucl. Phys.}\ }\textbf {\bibinfo {volume} {51}},\
  \bibinfo {pages} {223} (\bibinfo {year} {2003})}\BibitemShut {NoStop}%
\bibitem [{\citenamefont {Aoyama}\ \emph {et~al.}(2006)\citenamefont {Aoyama},
  \citenamefont {Myo}, \citenamefont {Katō},\ and\ \citenamefont
  {Ikeda}}]{10.1143/PTP.116.1}%
  \BibitemOpen
  \bibfield  {author} {\bibinfo {author} {\bibfnamefont {S.}~\bibnamefont
  {Aoyama}}, \bibinfo {author} {\bibfnamefont {T.}~\bibnamefont {Myo}},
  \bibinfo {author} {\bibfnamefont {K.}~\bibnamefont {Katō}}, \ and\ \bibinfo
  {author} {\bibfnamefont {K.}~\bibnamefont {Ikeda}},\ }\href {\doibase
  10.1143/PTP.116.1} {\bibfield  {journal} {\bibinfo  {journal} {Progress of
  Theoretical Physics}\ }\textbf {\bibinfo {volume} {116}},\ \bibinfo {pages}
  {1} (\bibinfo {year} {2006})},\ \Eprint
  {http://arxiv.org/abs/https://academic.oup.com/ptp/article-pdf/116/1/1/19571980/116-1.pdf}
  {https://academic.oup.com/ptp/article-pdf/116/1/1/19571980/116-1.pdf}
  \BibitemShut {NoStop}%
\bibitem [{\citenamefont {Aaij}\ \emph
  {et~al.}(2021{\natexlab{b}})\citenamefont {Aaij} \emph
  {et~al.}}]{LHCb:2020jpq}%
  \BibitemOpen
  \bibfield  {author} {\bibinfo {author} {\bibfnamefont {R.}~\bibnamefont
  {Aaij}} \emph {et~al.} (\bibinfo {collaboration} {LHCb}),\ }\href {\doibase
  10.1016/j.scib.2021.02.030} {\bibfield  {journal} {\bibinfo  {journal} {Sci.
  Bull.}\ }\textbf {\bibinfo {volume} {66}},\ \bibinfo {pages} {1278} (\bibinfo
  {year} {2021}{\natexlab{b}})},\ \Eprint {http://arxiv.org/abs/2012.10380}
  {arXiv:2012.10380 [hep-ex]} \BibitemShut {NoStop}%
\bibitem [{\citenamefont {Manohar}\ and\ \citenamefont
  {Wise}(2000)}]{Manohar:2000dt}%
  \BibitemOpen
  \bibfield  {author} {\bibinfo {author} {\bibfnamefont {A.~V.}\ \bibnamefont
  {Manohar}}\ and\ \bibinfo {author} {\bibfnamefont {M.~B.}\ \bibnamefont
  {Wise}},\ }\href@noop {} {\emph {\bibinfo {title} {{Heavy quark physics}}}},\
  Vol.~\bibinfo {volume} {10}\ (\bibinfo {year} {2000})\BibitemShut {NoStop}%
\bibitem [{\citenamefont {Liu}\ and\ \citenamefont {Oka}(2012)}]{Liu:2011xc}%
  \BibitemOpen
  \bibfield  {author} {\bibinfo {author} {\bibfnamefont {Y.-R.}\ \bibnamefont
  {Liu}}\ and\ \bibinfo {author} {\bibfnamefont {M.}~\bibnamefont {Oka}},\
  }\href {\doibase 10.1103/PhysRevD.85.014015} {\bibfield  {journal} {\bibinfo
  {journal} {Phys. Rev. D}\ }\textbf {\bibinfo {volume} {85}},\ \bibinfo
  {pages} {014015} (\bibinfo {year} {2012})},\ \Eprint
  {http://arxiv.org/abs/1103.4624} {arXiv:1103.4624 [hep-ph]} \BibitemShut
  {NoStop}%
\end{thebibliography}%


%

\end{document}